\documentclass{article}
\usepackage{spconf,amsmath,graphicx}

\usepackage{amssymb}
\usepackage{amsthm}
\usepackage{mathtools}
\usepackage{multirow}
\usepackage{cite}
\usepackage{array}
\usepackage{comment}
\usepackage{color}

\DeclareMathOperator*{\argmin}{argmin}

\title{HOW DOES END-TO-END SPEECH RECOGNITION TRAINING\\IMPACT SPEECH ENHANCEMENT ARTIFACTS?}
\name{\shortstack{Kazuma Iwamoto$^{1}$, Tsubasa Ochiai$^{2}$, Marc Delcroix$^{2}$,\\ \textit{Rintaro Ikeshita}$^{2}$, \textit{Hiroshi Sato}$^{2}$, \textit{Shoko Araki}$^{2}$, \textit{Shigeru Katagiri}$^{1}$}}
\address{
  $^1$Doshisha University
  $^2$NTT Corporation}
\begin{document}

\ninept
\maketitle
\begin{abstract}
Jointly training a speech enhancement (SE) front-end and an automatic speech recognition (ASR) back-end has been investigated as a way to mitigate the influence of \emph{processing distortion} generated by single-channel SE on ASR. In this paper, we investigate the effect of such joint training on the signal-level characteristics of the enhanced signals from the viewpoint of the decomposed noise and artifact errors. The experimental analyses provide two novel findings: 1) ASR-level training of the SE front-end reduces the artifact errors while increasing the noise errors, and 2) simply interpolating the enhanced and observed signals, which achieves a similar effect of reducing artifacts and increasing noise, improves ASR performance without jointly modifying the SE and ASR modules, even for a strong ASR back-end using a WavLM feature extractor. Our findings provide a better understanding of the effect of joint training and a novel insight for designing an ASR agnostic SE front-end.
\end{abstract}
\section{Introduction}

With the advent of deep learning, the performance of single-channel speech enhancement (SE) has greatly improved \cite{wang2018supervised} in terms of widely used SE evaluation measures such as signal-to-distortion ratio (SDR) \cite{vincent2006performance}, short-time objective intelligibility (STOI) \cite{taal2011an}, and perceptual evaluation of speech quality (PESQ) \cite{rix2001perceptual}.
However, automatic speech recognition (ASR) performance of the ASR systems (e.g., trained with standard multi-condition training strategy) tends to improve only slightly or even degrade when using a single-channel SE front-end compared to inputting observed noisy signals directly \cite{yoshioka2015ntt,chen2018building,fujimoto2019one,menne2019investigation}.
In this paper, we explore the cause of this phenomenon.

It is often assumed that the cause of such ASR performance limitations is \emph{processing distortion} induced by the single-channel SE systems.
To mitigate the processing distortion issue, joint training of the SE front-end and ASR back-end (e.g., fine-tuning the SE front-end with the ASR-level training objective) has been investigated, e.g., \cite{menne2019investigation,chang2022end-to-end,woo2020end,von2020multi,wang2016joint,wu2017end}.
Although the joint training scheme successfully improves ASR performance, it requires modifying the SE front-end and/or the ASR back-end, which is not always an option in practice, e.g., due to the need to use an already deployed system.
Moreover, joint training tends to degrade SE evaluation metrics such as SDR \cite{woo2020end,von2020multi}.
There has been no detailed analysis or interpretation of how such joint training schemes affect the signal-level characteristics of the enhanced signals generated by SE front-ends, although this could provide valuable insights for designing SE front-ends for ASR.

Vincent et al. \cite{vincent2006performance} introduced a technique to measure SE errors using orthogonal projection-based decomposition of the SE errors.
In our recent study \cite{iwamoto2022how}, we used this projection to identify the negative effect of decomposed artifact errors on ASR performance.
Here, we use a mathematical definition of artifact errors as the SE error projected onto the subspace orthogonal to the speech-noise subspace spanned by the speech and noise signals \cite{vincent2006performance}.
Moreover, we mathematically showed that observation adding (OA) post-processing, which simply interpolates the enhanced and observed signals without requiring the modification of the SE front-end or ASR back-end, reduces the artifact errors while increasing the noise errors, resulting in the ASR performance improvement.

Although the previous study \cite{iwamoto2022how} systematically proved the hypothesis that artifact errors have a more negative impact on ASR performance compared to remaining noise errors, the experimental setup \cite{iwamoto2022how} was limited, e.g., such negative effect was only validated with the conventional hybrid ASR system \cite{hinton2012deep}, whose SE and ASR modules were independently trained with a limited amount of speech and noise data.
Thus, it has not been fully revealed that such a hypothesis holds even for the recent state-of-the-art ASR systems \cite{chang2022end-to-end} that employs the recent end-to-end Transformer ASR model \cite{karita2019improving} and self-supervised learning (SSL)-based feature extraction model (WavLM \cite{chen2022wavlm}). Indeed, SSL models are trained with a huge amount of unlabeled speech and noise data, allowing the ASR back-end to be more robust to various speech and noise signals and could also potentially be more robust to artifacts.

In this paper, to consolidate the reliability of the findings in \cite{iwamoto2022how}, we analyze the negative effect of the artifact errors based on the state-of-the-art end-to-end ASR system with the SSL-based feature extractor. 
In addition, joint training of the SE front-end generates the enhanced signals optimal for ASR.
Therefore, this paper analyzes the effect of joint training from the viewpoint of decomposed noise and artifact errors.
This allows us to identify the properties of enhanced signals desirable for ASR more directly.
Through the analyses in this paper, we reveal three important findings:
\begin{enumerate}
\item
We experimentally demonstrate that training the SE front-end jointly with the ASR back-end improves ASR performance by reducing artifact errors at the cost of increasing noise errors on the enhanced signals generated by SE front-ends. This provides additional evidence for the negative effect of artifact errors on ASR performance.
\item
Artifacts errors severely impact the ASR performance even for state-of-the-art ASR back-ends using SSL-based feature extractor trained with a huge amount of speech and noise data.
These results suggest that appropriately dealing with the artifact errors for designing SE front-ends is essential for building robust ASR systems.
\item
The extremely simple OA post-processing, which does not require modifying the SE and ASR modules, achieves significant ASR performance improvement that is only slightly worse than the optimal but significantly more complex joint training scheme.
\end{enumerate}
We believe that these findings confirm that artifact errors play a detrimental effect on ASR and open novel research directions for single-channel robust ASR without relatively complex joint training schemes.



\section{Robust ASR pipeline with SE front-end}

In this section, we overview the robust ASR pipeline and its training procedures, which are used in the experimental evaluations.

\subsection{Processing pipeline}

\begin{figure}[t]
  \centering
  \includegraphics[width=0.8\linewidth]{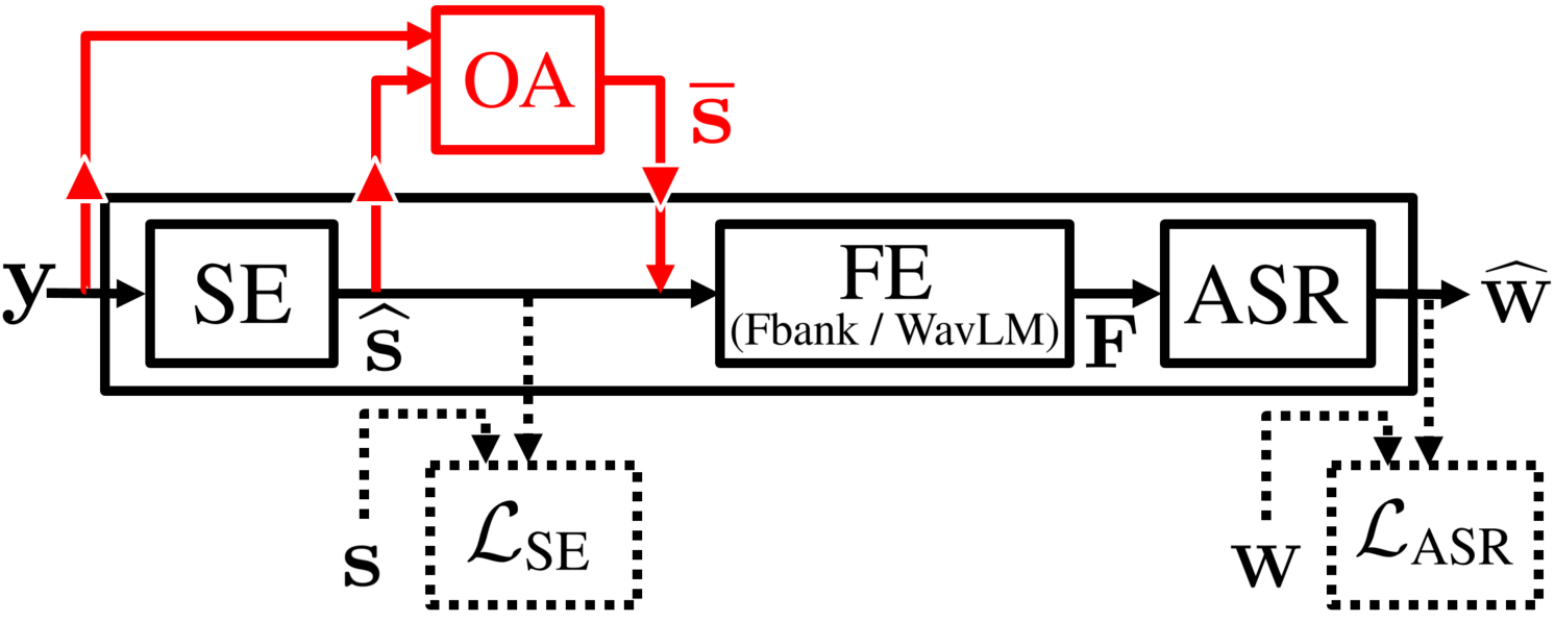}
\vspace{-1mm}
  \caption{Overview of robust ASR pipeline with SE front-end.}\label{fig:flowchart}
\vspace{-3mm}
\end{figure}

Figure~\ref{fig:flowchart} illustrates the processing pipeline of the evaluated robust ASR system that integrates SE, feature extraction (FE), and ASR modules, which we call the \emph{SE-FE-ASR} model.
Here, $\mathbf{y} \in \mathbb{R}^{\mathcal{T}}$ denotes a $\mathcal{T}$-length observed noisy signal in the time-domain.
$\widehat{\mathbf{W}} = \{ w_{n} \in \mathcal{V} \mid n = 1, \ldots, N \}$ denotes an $N$-length sequence of output labels, where $w_{n}$ represents an output symbol (e.g., character or byte pair encoding \cite{shibata1999byte}) at output time step $n$ and $\mathcal{V}$ is the vocabulary set.

In this paper, we follow the SE-FE-ASR model architecture used in a recent study \cite{chang2022end-to-end}.
We describe each of these modules below.

\textbf{SE module:}
First, the SE module generates the enhanced signal $\widehat{\mathbf{s}}  \in \mathbb{R}^{\mathcal{T}}$ given the observed noisy signal $\mathbf{y}$ as input:
\begin{align}
    \widehat{\mathbf{s}} = \text{SE}(\mathbf{y}; \boldsymbol{\theta}_{\text{SE}}),
    \label{eq:se}
\end{align}
where $\text{SE}(\cdot)$ represents the functional representation of the SE module with model parameters $\boldsymbol{\theta}_{\text{SE}}$.
The SE module suppresses the undesired noise included in the observed noisy signal $\mathbf{y}$.
We adopt the time-domain convolutional network (TDCN) \cite{luo2019conv} as SE module.

\textbf{FE module:}
Then, the FE module transforms the enhanced signal $\widehat{\mathbf{s}}$ to a $T$-length sequence of extracted features $\mathbf{F} = \{ \mathbf{f}_{t} \in \mathbb{R}^{D} \mid t=1, \ldots, T \}$ as:
\begin{align}
    &\mathbf{F} = \text{FE}(\widehat{\mathbf{s}}; \boldsymbol{\theta}_{\text{FE}}), \label{eq:featureextraction}
\end{align}
where $\text{FE}(\cdot)$ represents the functional representation of the FE module with parameters $\boldsymbol{\theta}_{\text{FE}}$.
Here, $\mathbf{f}_{t} \in \mathbb{R}^{D}$ denotes a $D$-dimensional intermediate feature vector at input time frame $t$.
We adopt two types of FE modules.
The first one extracts log Mel-Filterbank coefficients (Fbank), which is a de facto standard for ASR.
The other one uses WavLM Large \cite{chen2022wavlm}, which is a state-of-the-art SSL-based feature extraction model trained using a huge amount of unlabeled speech and noise data.

\textbf{ASR module:}
Finally, the ASR module generates the sequence of output labels $\widehat{\mathbf{W}}$ given the sequence of extracted features $\mathbf{F}$ as:
\begin{align}
    &\widehat{\mathbf{W}} = \text{ASR}(\mathbf{F}; \boldsymbol{\theta}_{\text{ASR}}), \label{eq:e2e-asr}
\end{align}
where $\text{ASR}(\cdot)$ represents the functional representation of the ASR module with parameters $\boldsymbol{\theta}_{\text{ASR}}$.
We adopt the joint CTC/attention-based encoder-decoder framework based on the Transformer architecture for the ASR module \cite{karita2019improving}.

\subsection{Joint training}
\label{sec:joint_training}

Previous studies \cite{chen2018building,woo2020end,von2020multi} have reported that when training independently the SE and ASR modules (e.g., trained with standard multi-condition training strategy), even an SE front-end that significantly improves the SE measures only slightly improves or even degrades ASR performance.
This behavior has been attributed to processing distortions generated by the SE front-end.
Several previous studies (e.g., \cite{menne2019investigation,chang2022end-to-end,woo2020end,von2020multi,wang2016joint,wu2017end}) proposed joint training or fine-tuning of the SE front-end and the ASR back-end to mitigate the effect of such processing distortions.

Let $\mathcal{L}_\text{SE}(\mathbf{s}, \widehat{\mathbf{s}}(\boldsymbol{\theta}_{\text{SE}}))$ denote the SE-level training objective (e.g., scale-invariant source-to-noise ratio (SI-SNR) \cite{luo2019conv}) defined from the reference and estimated signals generated by the SE module, denoted by $\mathbf{s}$ and $\widehat{\mathbf{s}}(\boldsymbol{\theta}_{\text{SE}})$, respectively.
Let $\mathcal{L}_\text{ASR}(\mathbf{W}, \widehat{\mathbf{W}}(\boldsymbol{\theta}_{\text{ASR}}))$ denote the ASR-level training objective (e.g., negative log-likelihood) defined from the reference and estimated transcriptions generated by the ASR module, denoted by $\mathbf{W}$ and $\widehat{\mathbf{W}}(\boldsymbol{\theta}_{\text{ASR}})$, respectively.
As shown in Figure~\ref{fig:flowchart}, the SE and ASR training objectives are computed on top of the SE and ASR modules, respectively.
The optimization procedure of the SE-FE-ASR model is described as:
\begin{align}
    \widehat{\boldsymbol{\Lambda}} = \argmin_{\boldsymbol{\Lambda}=\{\boldsymbol{\theta}_{\text{SE}}, \boldsymbol{\theta}_{\text{ASR}}\}} \mathcal{L}_\text{SE}(\mathbf{s}, \widehat{\mathbf{s}}(\boldsymbol{\theta}_{\text{SE}})) + \mathcal{L}_\text{ASR}(\mathbf{W}, \widehat{\mathbf{W}}(\boldsymbol{\theta}_{\text{ASR}})),
\end{align}
where $\boldsymbol{\Lambda}$ denotes the model parameters to be optimized.
The model parameters of the FE module $\boldsymbol{\theta}_{\text{FE}}$ could also be optimized in theory, but in this paper, we fix it by following the prior work \cite{chang2022end-to-end}.

There are three types of joint training (fine-tuning) schemes: 1) fine-tuning the SE front-end with the ASR-level training objective (i.e., set $\boldsymbol{\Lambda} = \{ \boldsymbol{\theta}_{\text{SE}} \}$), 2) fine-tuning the ASR back-end with the enhanced signals generated by the SE front-end (i.e., set $\boldsymbol{\Lambda} = \{ \boldsymbol{\theta}_{\text{ASR}} \}$), and 3) jointly training the SE and ASR modules (i.e., set $\boldsymbol{\Lambda} = \{ \boldsymbol{\theta}_{\text{SE}}, \boldsymbol{\theta}_{\text{ASR}} \}$).
Following a previous study \cite{chang2022end-to-end}, we first pre-train each module independently, using a SE loss $\mathcal{L}_{\text{SE}}$ \cite{luo2019conv} for the SE module, an SSL loss $\mathcal{L}_{\text{SSL}}$ \cite{chen2022wavlm} for the FE model, and an ASR loss $\mathcal{L}_{\text{ASR}}$ \cite{karita2019improving} for the ASR module.

Although such joint training schemes would be effective in improving ASR performance, it is not always possible 1) to tune the SE front-end for a specific ASR back-end and 2) to tune the ASR back-end for a specific SE front-end, e.g., due to the need to use already deployed SE/ASR modules or the cost of training and maintaining SE/ASR modules for different application scenarios.
Therefore, this paper evaluates both separate and joint training schemes.

\vspace{-1mm}
\section{Decomposed noise and artifact errors}

In this section, we review the definition of decomposed noise and artifact errors and introduce OA post-processing \cite{iwamoto2022how}.

\subsection{Orthogonal projection-based error decomposition}
\label{sec:error_decomposition}

The SE module inevitably produces estimation errors in the enhanced signal.
Based on the orthogonal projection, the total estimation error, i.e., $\mathbf{e}_{\text{total}} = \widehat{\mathbf{s}} - \mathbf{s}_{\text{target}}$, is decomposed into two types of error components \cite{vincent2006performance} as:
\begin{align}
    \mathbf{e}_{\text{total}} = \mathbf{e}_{\text{noise}} + \mathbf{e}_{\text{artif}},
    \label{eq:decomp}
\end{align}
where $\mathbf{s}_{\text{target}} \in \mathbb{R}^{\mathcal{T}}$ is the target source, $\mathbf{e}_{\text{noise}} \in \mathbb{R}^{\mathcal{T}}$ is the noise error, and $\mathbf{e}_{\text{artif}} \in \mathbb{R}^{\mathcal{T}}$ is the artifact error component.

These three components are derived with the orthogonal projections onto the subspace spanned by the reference source $\mathbf{s} \in \mathbb{R}^{\mathcal{T}}$ and noise $\mathbf{n} \in \mathbb{R}^{\mathcal{T}}$ signals as basis vectors.
We can obtain the projected signals as $\mathbf{s}_{\text{target}}=\mathbf{P}_{\mathbf{s}} \widehat{\mathbf{s}}$, $\mathbf{e}_{\text{noise}}=\mathbf{P}_{\mathbf{s},\mathbf{n}} \widehat{\mathbf{s}} - \mathbf{P}_{\mathbf{s}} \widehat{\mathbf{s}}$, and $\mathbf{e}_{\text{artif}}=\widehat{\mathbf{s}} - \mathbf{P}_{\mathbf{s},\mathbf{n}} \widehat{\mathbf{s}}$, where $\mathbf{P}_{\mathbf{s}} \in \mathbb{R}^{\mathcal{T} \times \mathcal{T}}$ is the orthogonal projection matrix onto the subspace spanned by $\mathbf{s}$, and $\mathbf{P}_{\mathbf{s},\mathbf{n}} \in \mathbb{R}^{\mathcal{T} \times \mathcal{T}}$ is that onto the subspace spanned by $\mathbf{s}$ and $\mathbf{n}$.

Based on the above decomposed components, three SE evaluation measures, i.e., 1) SDR, 2) signal-to-noise ratio (SNR), and 3) signal-to-artifact ratio (SAR), are defined \cite{vincent2006performance} as:
\begin{align}
    \text{SDR} &\coloneqq 10 \log_{10} (\lvert \lvert \mathbf{s}_{\text{target}} \rvert \rvert^{2} / \lvert \lvert \mathbf{e}_{\text{noise}} + \mathbf{e}_{\text{artif}} \rvert \rvert^{2}), \\
    \text{SNR} &\coloneqq 10 \log_{10} (\lvert \lvert \mathbf{s}_{\text{target}} \rvert \rvert^{2} / \lvert \lvert \mathbf{e}_{\text{noise}} \rvert \rvert^{2}), \\
    \text{SAR} &\coloneqq 10 \log_{10} (\lvert \lvert \mathbf{s}_{\text{target}} + \mathbf{e}_{\text{noise}} \rvert \rvert^{2} / \lvert \lvert \mathbf{e}_{\text{artif}} \rvert \rvert^{2}). \label{eq:sar}
\end{align}
$\text{SDR}$ measures the impact of total error, while $\text{SNR}$ and $\text{SAR}$ measure the impact of each error separately.

We compare the SNR and SAR scores before and after joint training to analyze the effect of joint training, as described in Section~\ref{sec:joint_training}, which has not been investigated in the prior works \cite{menne2019investigation,chang2022end-to-end,woo2020end,von2020multi,wang2016joint,wu2017end}.

\subsection{Artifact error compensation with observation adding}

Recent studies \cite{iwamoto2022how,zorila2022speaker} proposed using the OA technique as the post-processing of the SE module.
The OA module is inserted between the SE and FE modules shown in Figure~\ref{fig:flowchart}.
The OA consists of simply interpolating the enhanced $\widehat{\mathbf{s}}$ and observed $\mathbf{y}$ signals as:
\begin{align}
    \overline{\mathbf{s}} = (1-\omega_{\text{obs}})\ \widehat{\mathbf{s}} + \omega_{\text{obs}}\ \mathbf{y},\label{eq:oa}
\end{align}
where $0.0 \leq \omega_{\text{obs}} \leq 1.0$ is a weight hyperparameter that controls the balance of the enhanced and observed signals.

In \cite{iwamoto2022how}, we mathematically proved that adding the observed signal increases the norm of $\mathbf{s}_{\text{target}}$ and $\mathbf{e}_{\text{noise}}$, without changing that of the artifacts $\mathbf{e}_{\text{artif}}$ in a mild condition.
We demonstrated that the artifact component is the most detrimental and confirmed experimentally that the OA effectively improved ASR performance of the independently trained SE and ASR pipeline.

In this paper, we compare the effectiveness of the simple OA-based SE front-end to the joint training-based one, which is directly optimized for improving ASR performance.

\section{Experiments}

\begin{table*}[t]
  \caption{SE and ASR results of evaluated systems (SDR, STOI, PESQ, SNR, SAR (higher is better) and WER (lower is better)).}
\vspace{2mm}
  \label{tab:result}
  \centering
  \renewcommand{\arraystretch}{1.0}
  \scalebox{0.9}{
  \begin{tabular}{@{} c | wc{0.85cm} | wc{0.4cm} wc{0.4cm} | wc{0.55cm} wc{0.65cm} wc{0.65cm} wc{0.55cm} wc{0.55cm} wc{0.55cm} wc{0.55cm} | wc{0.55cm} wc{0.55cm} wc{0.55cm} wc{0.65cm} wc{0.65cm} wc{0.55cm} wc{0.55cm} wc{0.55cm} wc{0.55cm} @{}}
    \hline
    & & & & \multicolumn{7}{c |}{SE-FE-ASR} & \multicolumn{7}{c}{SE-OA-FE-ASR} \\
    & & & & \multicolumn{7}{c |}{} & \multicolumn{7}{c}{} \vspace{-3mm}\\
    & FE & \shortstack{FT\\SE} & \shortstack{FT\\ASR} & \shortstack{SDR$\uparrow$\\(simu)} & \shortstack{STOI$\uparrow$\\(simu)} & \shortstack{PESQ$\uparrow$\\(simu)} & \shortstack{SNR$\uparrow$\\(simu)} & \shortstack{SAR$\uparrow$\\(simu)} & \shortstack{WER$\downarrow$\\(simu)} & \shortstack{WER$\downarrow$\\(real)} & \shortstack{$\omega_{\text{obs}}$\\(simu)} & \shortstack{$\omega_{\text{obs}}$\\(real)} & \shortstack{SDR$\uparrow$\\(simu)}
    & \shortstack{STOI$\uparrow$\\(simu)} & \shortstack{PESQ$\uparrow$\\(simu)}& \shortstack{SNR$\uparrow$\\(simu)} & \shortstack{SAR$\uparrow$\\(simu)} & \shortstack{WER$\downarrow$\\(simu)} & \shortstack{WER$\downarrow$\\(real)} \\
    \hline
     (1) & Fbank & $\times$ & -- & 3.6 & 0.81 & 1.20 & 4.7 & -- & 18.0 & 16.4 & -- & -- & -- & -- & -- & -- & -- & -- & -- \\
     (2) & & -- & -- & 9.0 & 0.88 & 1.56 & 21.5 & 9.4 & 27.4 & 27.1 & 0.6 & 0.8 & 6.9 & 0.86 & 1.37 & 9.3 & 14.5 & 16.8 & 15.8 \\
     (3) & & -- & \checkmark & 9.0 & 0.88 & 1.56 & 21.5 & 9.4 & 19.5 & 19.2 & 0.1 & 0.3 & 9.1 & 0.89 & 1.71 & 18.3 & 10.0 & 18.7 & 17.5 \\
     (4) & & \checkmark & -- & 7.6 & 0.85 & 1.31 & 12.7 & 10.5 & 17.2 & 15.1 & 0.3 & 0.1 & 7.0 & 0.85 & 1.29 & 10.0 & 12.3 & 16.6 & 15.1 \\
     (5) & & \checkmark & \checkmark & 8.6 & 0.88 & 1.43 & 15.5 & 10.2 & 16.8 & 15.1 & 0.1 & 0.3 & 8.5 & 0.88 & 1.42 & 14.1 & 10.8 & 16.8 & 15.3 \\
     \hline
     (6) & WavLM & $\times$ & -- & 3.6 & 0.81 & 1.20 & 4.7 & -- & 8.4 & 4.8 & -- & -- & -- & -- & -- & -- & -- & -- & -- \\
     (7) & & -- & -- & 9.0 & 0.88 & 1.56 & 21.5 & 9.4 & 12.3 & 9.4 & 0.5 & 0.6 & 7.6 & 0.87 & 1.42 & 10.6 & 13.3 & 6.4 & 4.3 \\
     (8) & & -- & \checkmark & 9.0 & 0.88 & 1.56 & 21.5 & 9.4 & 11.4 & 8.8  & 0.4 & 0.5 & 8.2 & 0.88 & 1.48 & 12.1 & 12.3 & 6.3 & 4.3 \\
     (9) & & \checkmark & -- & 9.0 & 0.88 & 1.48 & 16.3 & 10.8 & 6.2 & 3.9 & 0.3 & 0.1 & 8.2 & 0.87 & 1.41 & 12.0 & 12.6 & 5.9 & 3.8 \\
     (10) & & \checkmark & \checkmark & 8.9 & 0.87 & 1.47 & 15.7 & 10.7 & 6.1 & 3.8 & 0.3 & 0.2 & 8.1 & 0.87 & 1.40 & 11.7 & 12.5 & 5.9 & 3.8  \\
    \hline
  \end{tabular}
  }
\vspace{-3mm}
\end{table*}

\subsection{Evaluated data}

We conducted the experiments on the CHiME-4 corpus \cite{barker2015third}, which is widely used in the noise-robust ASR community.
The corpus contains not only simulated data but also real recordings from noisy public environments including bus, cafe, pedestrian and street.
The training set contains 15 hours of simulated utterances spoken by 83 speakers and three hours of real utterances spoken by four speakers.
The development/evaluation sets contain 1,640/1,320 simulated and real utterances spoken by four speakers, respectively.

\subsection{Evaluated systems}

We followed the SE-FE-ASR model architecture used in \cite{chang2022end-to-end}, which achieves a state-of-the-art ASR performance for single-channel setup on the CHiME-4 task.
We implemented it based on the CHiME-4 recipe of the ESPnet toolkit \cite{watanabe2018espnet} and used the same network architectures and hyperparameter settings of \cite{chang2022end-to-end}.

\subsubsection{Separate pre-training of SE, FE, and ASR modules}

\ 
\vspace{-5mm}

\textbf{SE module:}
As the SE module's architecture, we adopted the TDCN model (i.e., Conv-TasNet \cite{luo2019conv}).
By following the notations of a previous study \cite{luo2019conv}, we set the hyper-parameters as follows: $N = 256$, $B = 256$, $H = 512$, $P = 3$, $X = 4$, and $R = 2$.
During the pre-training stage, we trained the SE module with the SI-SNR loss \cite{luo2019conv} by adopting the Adam optimizer \cite{kingma2015adam} with an initial learning rate of 0.001.

\textbf{FE module:}
As the FE module’s architecture, we adopted the conventional Fbank and the recent WavLM Large model.
For the Fbank, we used a Hanning window with a length and shift set at 25 ms and 10 ms, respectively.
For the WavLM, the weighted sum of the output of different layers of the WavLM model is used as the input for the ASR back-end \cite{yang2021superb}, where the weight parameters are also optimized simultaneously during the ASR pre-training.
Following a previous study \cite{chang2022end-to-end}, During the fine-tuning stage of the SE and ASR modules, the FE modules (i.e., Fbank and WavLM parameters) are not updated.

\textbf{ASR module:}
We adopted the joint CTC/attention-based encoder-decoder model with 12 layers for the encoder and six layers for the decoder.
For each Transformer layer, the number of attention heads was four, and the dimension of the linear projection was 2,048.
The encoder included two convolutional layers that downsample the input feature sequence and increase the frame shift from 10 ms to 40 ms.
During the pre-training stage, we trained the ASR module with the CTC/Attention negative log likelihood loss by adopting the Adam optimizer with a peak learning rate of 0.001 and 20,000 steps to warm up.
During the beam search decoding, we used a character-level Transformer language model \cite{irie2019language} with a language model weight of 1.0.

\vspace{-2mm}
\subsubsection{Joint fine-tuning of SE and ASR modules}

We evaluated four SE-FE-ASR systems obtained by 1) separately training the SE and ASR modules (i.e., no fine-tuning), 2) re-training only the SE module (i.e., SE fine-tuning), 3) re-training only the ASR module (i.e., ASR fine-tuning), and 4) jointly training the SE and ASR modules (i.e., overall fine-tuning).
During the fine-tuning stage, we trained the SE and ASR modules as described in Section~\ref{sec:joint_training} by adopting the Adam optimizer with an initial learning rate of 0.0005.

\vspace{-2mm}
\subsection{Results}

Table~\ref{tab:result} shows the SE (i.e., SDR, STOI, PESQ, SNR, SAR) and ASR (i.e., word error rate (WER)) performance measures.
We show the results for 1) the standard SE-FE-ASR models (left side) and 2) the SE-FE-ASR model with OA post-processing (right side), which we call the \emph{SE-OA-FE-ASR} model.
The ``FT SE'' and ``FT ASR'' columns indicate whether the SE and ASR modules are fine-tuned.
``$\times$'' in the ``FT SE'' column shows the results without any SE front-end, i.e., the SE and ASR scores of the noisy baseline.
Interestingly, as in \cite{chang2022end-to-end}, the simulated data are more challenging than the real data for this setup.

\vspace{-2mm}
\subsubsection{Evaluating separately pre-trained SE-FE-ASR models}

First, we look at the results of the SE-FE-ASR models (left side of Table~\ref{tab:result}).
From the table, we confirm that the SE-FE-ASR models with separate training (i.e., systems (2) and (7)) do not improve ASR performance compared to using the observed noisy signal directly (i.e., systems (1) and (6)), even when using strong SSL-based ASR back-end.
For example, the WER on the real evaluation data (i.e., WER (real)) degrades from 4.8 \% to 9.4\% when using an SE front-end with a WavLM-based ASR back-end (i.e., systems (6) and (7)).
This is similar to what has been reported in previous studies \cite{chen2018building,iwamoto2022how}.
This result suggests that the ASR back-ends also suffer from the processing distortion issues of the SE front-ends, even with recent state-of-the-art transformer-based models with an SSL feature extractor (i.e., WavLM Large) trained using a huge amount of speech and noise data.

\vspace{-2mm}
\subsubsection{Evaluating effect of SE and ASR fine-tuning}
\label{section:experiment1}

We then evaluated the effect of fine-tuning the SE and ASR modules.
From the table, we confirmed that even when the ASR modules are fine-tuned on the enhanced signals (i.e., systems (3) and (8)), the SE modules still degrade the ASR performance compared to using the observed noisy signals directly (i.e., systems (1) and (6)).

On the other hand, fine-tuning the SE modules with the ASR-level training objective (i.e., systems (4) and (9)) improves WER relatively by, e.g., 18.8 \% (from 4.8 \% to 3.9 \%) for WER (real) with system (9).
This result suggests that fine-tuning the SE module is an effective way to mitigate the processing distortion issue.
However, looking at the widely used SE measures such as SDR, STOI, and PESQ, we do not observe significant improvement, and even sometimes degradation.
Interestingly, by performing the error decomposition in Section~\ref{sec:error_decomposition}, we observe that fine-tuning the SE front-end reduces artifacts (increases SAR) while increasing noise (decreases SNR).
A similar trend is also observed when fine-tuning both the SE and ASR models (i.e., systems (5) and (10)).
Consequently, we can conclude that fine-tuning may not improve SE performance in general but does reduce artifacts harmful to ASR by allowing less noise reduction.
These results strongly support the findings of our prior work \cite{iwamoto2022how} that the artifact errors have a more negative impact on ASR performance, because they revealed that the enhanced signal optimal for ASR generated by joint training has lower level of artifact errors.

\vspace{-2mm}
\subsubsection{Evaluating effect of observation adding post-processing}
\label{section:experiment2}

Finally, we look at the results of the SE-OA-FE-ASR models with the OA post-processing (right side of Table~\ref{tab:result}) and evaluate the effectiveness of the OA post-processing and the necessity of the joint training scheme of the SE and ASR modules.
We tune the weight hyperparameter $\omega_{\text{obs}}$ for each system by choosing the value in the range $\omega_{\text{obs}} = \{ 0.0, 0.1, \ldots, 1.0 \}$ that achieves the best WER on the development set.
We show in the table the tuned weight values.
As expected, when fine-tuning the SE/ASR modules, lower weights tended to give better performance.

From the table, comparing the SNR and SAR scores with and without OA (i.e., SE-FE-ASR and SE-OA-FE-ASR), we confirm that the OA post-processing reduces the artifact errors while increasing the noise errors.
This effect successfully improves the ASR performance even with a strong ASR back-end with an SSL-based feature extractor.
The effect of OA is particularly strong for separately trained SE and ASR modules (i.e., SE-OA-FE-ASR (2) and (7)) and when fine-tuning only the ASR module (i.e., SE-OA-FE-ASR (3) and (8)).
The OA has a similar effect to that of fine-tuning the SE module with the ASR training objective (i.e., SE-FE-ASR (4) and (9)), although their SNR and SAR scores showed different balances.
These results suggest that the simple OA post-processing produces a similar effect to that of a more complex joint training scheme.
Therefore, it is promising as a practical alternative in cases where modifying the SE or ASR modules is not an option, e.g., when relying on already deployed independent SE and ASR systems.

On the other hand, by comparing systems (7) and (9) that use the same ASR back-end, we observe that system (7) without fine-tuning but using OA (i.e., SE-OA-FE-ASR (7)) achieves relative WER performance improvements of 23.8 \% and 10.4 \% for simu and real sets, respectively, while system (9) without OA but using fine-tuning (i.e., SE-FE-ASR (9)) achieves relative improvements of 26.2 \% and 18.8 \%.
We performed the matched pairs sentence-segment word error (MAPSSWE) significance test \cite{gillick1989some}.
Interestingly, the performance difference between OA and joint training (i.e., SE-OA-FE-ASR (7) and SE-FE-ASR (9)) is statistically significant ($p < 0.01$) for the real dataset but not for the simu dataset.
Besides, we confirmed that both approaches improved WER over the noisy baseline and that the improvement is statistically significant ($p < 0.01$).
These results show that OA still has room for further improvement of WER performance compared to joint training, which would be the upper bound of the SE front-end for ASR, but it can achieve significant performance improvement.

\vspace{-1mm}
\section{Conclusions}

This paper analyzed the effect of SE and ASR joint training schemes on the signal-level characteristics of the enhanced signals generated by SE front-ends.
Through experimental evaluations, this paper reached two novel findings: 1) ASR-level training of the SE front-end reduces artifact errors while increasing noise errors, and 2) the SE front-end with simple OA post-processing achieves the ASR performance improvement without jointly modifying the SE and ASR modules, even for the state-of-the-art ASR models using SSL models trained with a huge amount of speech and noise data.
 Future works include investigations of training schemes for designing an ASR agnostic SE front-end beyond that with the OA post-processing by, for example, developing an SE front-end that reduces the artifact errors while maintaining the noise errors to the extent possible.

\vfill\pagebreak

\bibliographystyle{IEEEtran}
\bibliography{mybib}

\end{document}